\documentclass[runningheads]{llncs}

\usepackage{tcdef}
\usepackage{graphicx}
\usepackage{caption}
\usepackage{amssymb}
\usepackage{amsmath}
\usepackage{color}
\usepackage{url}
\usepackage{enumitem}
\usepackage{framed,color}
\definecolor{shadecolor}{rgb}{.95,.95,.95}

\newcommand{\myskip}[1]{}

\begin{document}

\mainmatter

\title{Domain Decomposition for Heterojunction Problems in Semiconductors}

\author{Timothy Costa$^\dagger$ \and David Foster$^\ddagger$ \and
Malgorzata Peszynska$^\dagger$}
\institute{$^\dagger$ Department of Mathematics,
Oregon State University, Corvallis, OR, 97330, USA \\
$^\ddagger$ Department of Physics, Oregon State University, Corvallis, OR, 97330, USA}
\maketitle
\begin{abstract}
We present a domain decomposition approach for the simulation of
charge transport in heterojunction semiconductors.
The problem is characterized by a large variation of primary
variables across an interface region of a size much smaller than
the device scale, and requires a multiscale approach in which that region is
modeled as an internal boundary. The model combines drift diffusion
equations on subdomains coupled by thermionic emission heterojunction
model on the interface which involves a nonhomogeneous jump computed
at fine scale with Density Functional Theory. Our full domain decomposition
approach extends our previous work for the potential equation only,
and we present perspectives on its HPC implementation. The model
can be used, e.g., for the design of higher efficiency solar cells
for which experimental results are not available. More generally,
our algorithm is naturally parallelizable and is
a new domain decomposition paradigm for problems with multiscale
phenomena associated with internal interfaces and/or boundary layers.
\keywords{Domain
  Decomposition, Drift-Diffusion Equations, Density Functional Theory,
  Heterojunction, Multiscale Semiconductor Modeling, Solar Cells.}
\end{abstract}

\section{Introduction}
\label{sec:intro}

In this paper we present a multiscale approach for heterojunction
interfaces in semiconductors, part of a larger
  interdisciplinary effort between computational mathematicians,
  physicists, and material scientists interested in building more
  efficient solar cells.  The higher efficiency (may) arise from
  putting together different semiconductor materials, i.e., creating a
  {\em heterojunction}.

The computational challenge is that phenomena at heterojunctions must
be resolved at the angstrom scale while the size of the device is on
the scale of microns, thus it is difficult to simultaneously account
for correct physics and keep the model computationally tractable.  To
model charge transport at the device scale we use the drift diffusion
(D-D) system \cite{markowich86}.  For interfaces, we follow the
approach from \cite{HorioYanai} in which the interface region is
shrunk to a low-dimensional internal boundary, and physics at this
interface is approximated by the thermionic emission model (TEM) which
consists of unusual internal boundary conditions with jumps.

We determine the data for these jumps from an angstrom scale
calculation using Density Functional Theory (DFT), and we model the
physics away from the interface by the usual (D-D) equations
coupled by TEM. The D-D model can be hard-coded as a
  monolithic approach which appears intractable and/or impractical in
  2d/3d with complicated interface geometries. Our proposed
alternative is to apply a domain decomposition (DDM) approach which
allows the use of ``black box'' D-D solvers in subdomains, and
enforces the TEM conditions at the level of the DDM
driver. DDM have been applied to D-D, e.g., in
  \cite{LinShadid10a,LinShadid10b}, where the focus was on a multicore
  HPC implementation of efficiently implemented suite of linear and
  nonlinear solvers. Here we align the DDM with handling microscale
  physics at material interfaces. More importantly, fully decoupling
the subdomains is a first step towards a true multiscale simulation
where the behavior in the heterojunction region is treated
simultaneously by a computational method at microscale.

The DDM approach we propose is non-standard because of the
nonhomogeneous jumps arising from TEM. In \cite{haii} we presented the
\textbf{DDP} algorithm for the potential equation. In this paper we
report on the next nontrivial step which involves carrier transport
equations. Here the interface model is an unusual Robin-like interface
equation. The algorithm \textbf{DDC} works well and has promising
properties.

This paper consists of the following. In Section~\ref{sec:mod} we
describe the model.  In Section~\ref{sec:ddalgs} we present our
domain decomposition algorithms, and in Section~\ref{sec:scsim}
we present numerical results for the simulation of two semiconductor
heterojunctions.  Finally in Section~\ref{sec:conc} we present
conclusions, HPC context, and describe future work.

\section{Computational Model for Coupled Scales}
\label{sec:mod}

The continuum D-D model with TEM is described first, followed
  by the angstrom scale DFT model.

\subsection{Device scale continuum models: drift diffusion (D-D) system}

Let $\Omega \in \mathbb{R}^N$, $N \in \{1,2,3\}$, be an open connected
set with a Lipschitz boundary $\partial \Omega$. Let $\Omega_i \in
\Omega$, $i = 1,2$, be two non-overlapping subsets of $\Omega$
s.t. $\overline{\Omega}_1 \cup \overline{\Omega}_2 =
\overline{\Omega}$, $\Omega_1 \cap \Omega_2 = \emptyset$, and denote
$\Gamma:=\overline{\Omega}_1 \cap \overline{\Omega}_2$.  We assume
$\Gamma$ is a $N$-$1$ dimensional manifold, and $\Gamma \cap \partial
\Omega = \emptyset$.  Each subdomain $\Omega_i$ corresponds to a
distinct semiconductor material, and $\Gamma$ the interface between
them.  We adopt the following usual notation: $w_i =
w|_{\Omega_i}$, $w_i^\Gamma = \lim_{x\to \Gamma} w_i$, and
$\left[w\right]_\Gamma = w_2^\Gamma - w_1^\Gamma$ denotes
the jump of $w$.

In the bulk semiconductor domains $\Omega_i$, $i=1,2$, the charge
transport is described by the D-D system: a potential equation solved
for electrostatic potential $\psi$, and two continuity equations
solved for the Slotboom variables $u$ and $v$; these relate to the
electron and hole densities $n$ and $p$, respectively, via $n =
\delta_n^2 e^\psi u, \; p = \delta_p^2 e^{-\psi} v$. (The scaling
parameters $\delta_n^2$ and $\delta_p^2$ depend on the material and
the doping profile).  We recall that in Slotboom variables the
continuity equations are self-adjoint \cite{markowich86}. The
stationary D-D model is
\ba
\label{eq:pot}
& & -\nabla \cdot (\epsilon_i \nabla \psi_i) = \frac{1}{\eta}
(\delta_p^2 e^{-\psi_i} v - \delta_n^2 e^{\psi_i} u + N_T)
:= q(\psi_i, p_i, n_i), \\
\label{eq:nsb}
& & -\nabla \cdot (D_{n_i} \delta_n^2 e^{\psi_i}\, \nabla u_i) = R(\psi_i, u_i, v_i), \\
\label{eq:psb}
 & & -\nabla \cdot (D_{p_i} \delta_p^2 e^{-\psi_i}\, \nabla v_i) = -R(\psi_i, u_i, v_i).
\ea
For background on the D-D model the reader is referred to
\cite{bank,jerome-sdbook,markowich86,markowich,seeger,selber}.
In \eqref{eq:pot}--\eqref{eq:psb} we use data: the net doping profile
$N_T$, a given expression for the electron-hole pair generation and
recombination $R$, electrical permittivity $\epsilon$, and electron and
hole diffusivities $D_n$, $D_p$. Also, $\eta$ is another scaling
parameter \cite{haii}.

\begin{figure}[ht]
\centering
\includegraphics[width=.49\textwidth]{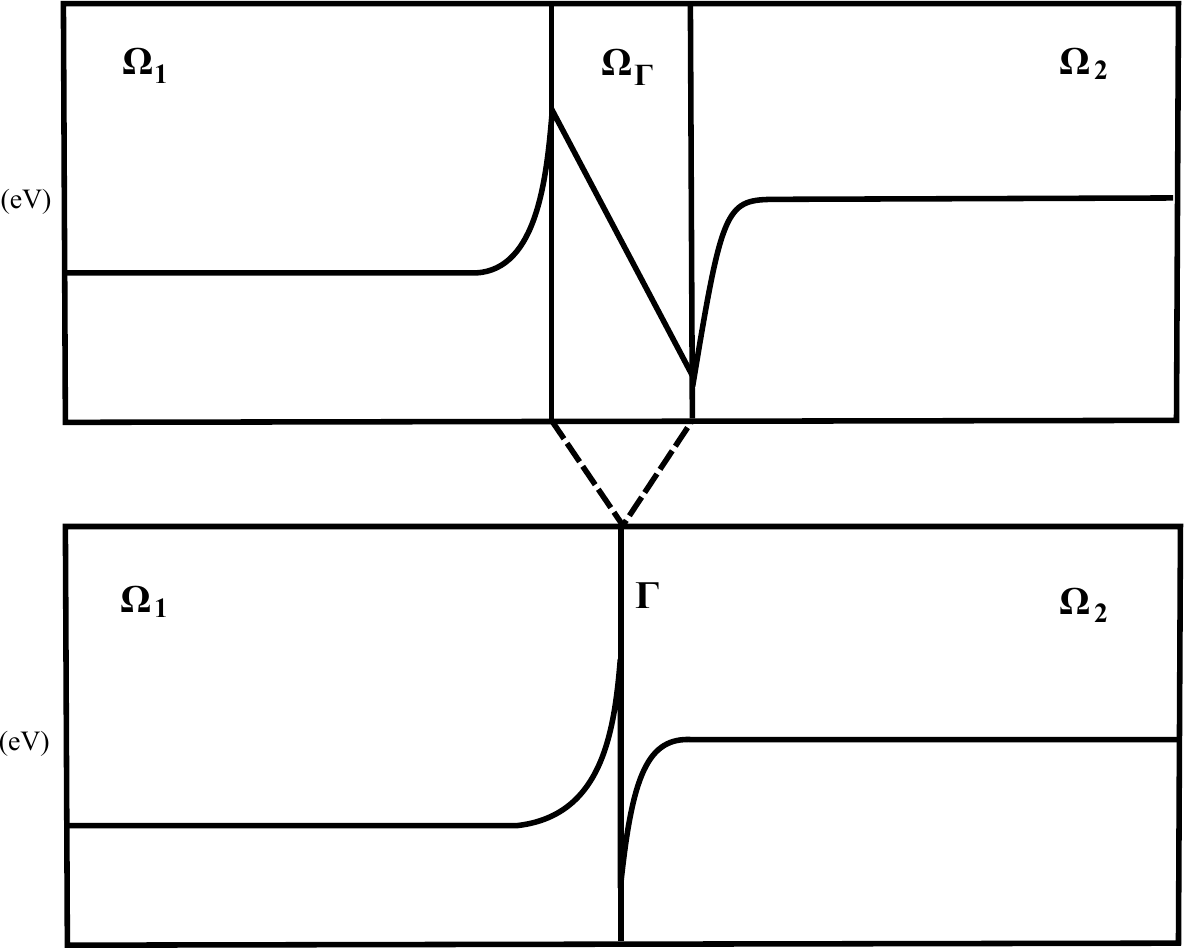}
\includegraphics[width=.49\textwidth]{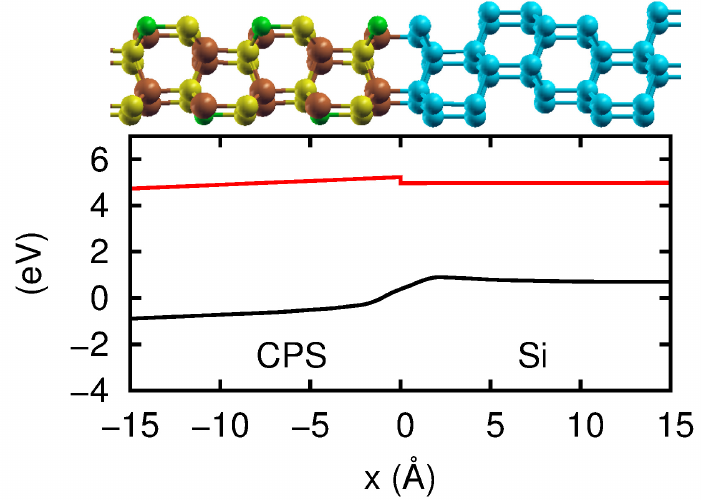}
\caption{\label{fig:supercell}Left top:
  schematic plot of potential across 1D interface region for Structure
  1 (actual simulation in Figure~\ref{fig:mondd}). Left bottom:
  schematic plot of potential with idealized heterojunction interface.
  Right top: interface atomic structure. Right bottom: smoothed
  local pseudopotential from the DFT calculation (black), and valence
  band jump construction (red), which determines
  $a_i^n$, $a_i^p$, and $\psi_\triangle$ \cite{haii}}
\end{figure}

The model \eqref{eq:pot}--\eqref{eq:psb} is completed with external
boundary conditions; we impose Dirichlet conditions for the potential
and recombination-velocity (Robin type) conditions for electron and
hole densities. To this we add the TEM transmission conditions on the
interface \cite{HorioYanai}
\ba
\label{eq:tranpot}
& &
\begin{array}{lr}
\left[\psi\right]_\Gamma = \psi_\triangle, &
\hspace{4cm} \left[\epsilon \dpd{\psi}{\nu} \right]_\Gamma = 0,
\end{array} \\
\label{eq:tranu}
& &
\begin{array}{lr}
J_{n_1} = a_2^n (e^{\psi} u)_2^\Gamma - a_1^n (e^\psi u)_1^\Gamma, &
\hspace{1.08cm} \left[J_n\right]_\Gamma = 0,
\end{array} \\
\label{eq:tranv}
& &
\begin{array}{lr}
J_{p_1} = a_1^p (e^{-\psi} v)_1^\Gamma - a_2^p (e^{-\psi} v)_2^\Gamma, &
\hspace{0.8cm} \left[J_p\right]_\Gamma = 0.
\end{array}
\ea
Here $J_n$ and $J_p$ are the electron and hole currents
\ba
J_n = D_n \delta_n^2 e^{\psi} \nabla u,
\\
J_p = D_p \delta_p^2 e^{-\psi} \nabla v.
\ea
Also, $a_i^n$ and $a_i^p$ are constants dependent on material
properties and temperature, and $\psi_\triangle$ is a jump
discontinuity in the electrostatic potential. These can be determined
by a DFT calculation, see Figure~\ref{fig:supercell} for illustration.

The model \eqref{eq:pot}--\eqref{eq:tranv} must be discretized. Here
we use simple finite difference formulation following
\cite{markowich86}, \cite{selber} with $N$ nodal unknowns; we
skip details for brevity.  In what follows we refer to $\psi,u,v,n,p$
meaning their discrete counterparts.

%%%%%%%%
\subsection{Density Functional Theory for Atomic Scale}
Heterojunction parameters $a_i^n,a_i^p,\psi_\triangle$ in
\eqref{eq:tranpot}--\eqref{eq:tranv} are determined by quantum
mechanics of electrons.
% TC removed:
%First principles methods solve a 3D quantum
%electron system having discrete atoms with a characteristic spacing of
%2-4 angstroms, see Fig.~\ref{fig:supercell} for illustration.
%
The Schr\"{o}dinger equation solved for wave function $\Psi$ is
fundamental for quantum behavior, but the problem of interacting
$N$ electrons is computationally intractable for large $N$.

DFT \cite{fiolhais,engel2011density} provides an efficient method of
determining material properties from first principles by shifting
focus from wave functions $\Psi$ to electron density,
$n(\boldsymbol{r})$.  The density sought in DFT is a function in
$\mathbb{R}^3$, while the Schr\"odinger equation is solved for $\Psi \in
\mathbb{C}^{3N}$.  Finding $n$ is possible via application of the
theory of Hohenberg and Kohn to a minimization problem in $n$, and is
equivalent to the solution of the Schr\"{o}dinger equation for the
ground state. However, an energy functional $F[n]$ needed in
the minimization principle in DFT is unknown, and DFT
requires approximations to $F[n]$.  The Kohn-Sham equations provide a
basis for these approximations, and their solution can be
found iteratively \cite{fiolhais,engel2011density}.

DFT is a widely used, low cost, first principles method which solves
the zero temperature, zero current ground state of a system
\cite{fiolhais,engel2011density}.  The local pseudopotential
calculated by DFT is continuous at an interface (see
Figure~\ref{fig:supercell}), and can be used with known material
properties to obtain the change in the continuous electrostatic
potential $\psi$ occurring close to a heterojunction.  \myskip{The
  transport, scattering, and recombination/generation processes
  involved in an active semiconductor device are not suitable for
  handling by first principles calculation methods, and remain in the
  realm of macroscopic (device scale) models.} The potential jump
(offset) $\psi_\triangle$ is a ground state property of the
heterojunction structure, and DFT solution in the heterojunction
region provides the data needed for TEM.

For the needs of this paper, we perform DFT calculations using the
VASP code \cite{kressel}, with exchange-correlation treated using the
Generalized Gradient Approximation and a $+U$ Hubbard term ($U=6$ eV)
for the Cu-$d$ orbitals \cite{haii,fiolhais,engel2011density}.

\myskip{Using 48 cpu cores, the primary DFT computation time was
20 hours.}

\section{Domain Decomposition for Continuum Model}
\label{sec:ddalgs}

The procedure to solve \eqref{eq:pot}--\eqref{eq:tranv} numerically
is a set of nested iterations, with three levels of nesting.

First, when solving \eqref{eq:pot}--\eqref{eq:tranv}, we employ the
usual Gummel Map \cite{quan-jer,jerome-sdbook}, an iterative
decoupling technique within which we solve each component equation of
\eqref{eq:pot}--\eqref{eq:psb} independently.  Note that each equation
is still nonlinear in its primary variable, thus we must use Newton's
iteration.

Furthermore, each component equation employs DDM independently to
resolve the corresponding part of TEM. In particular, we solve
potential equation \eqref{eq:pot} with \eqref{eq:tranpot}, the
electron transport \eqref{eq:nsb} with \eqref{eq:tranu}, and the hole
transport \eqref{eq:psb} with \eqref{eq:tranv}.  The DDM we use is an
iterative substructuring method designed as a Richardson scheme
\cite{qvddbook} to resolve the TEM, defined and executed independently
for each component.  In what follows $\theta > 0$ is an acceleration
parameter, different for each component equation. Since the DDM
algorithm for $p$ equation is entirely analogous to that for $n$
equation, we only define the latter.

Last, each subdomain solve of the DDM is nonlinear, and we use
Newton-Raphson iteration to resolve this.

\subsection{Domain Decomposition for Potential
Equation \eqref{eq:pot}, \eqref{eq:tranpot}}
\label{sub:potalg}

Here we seek the interface value of $\lambda$ with which
\eqref{eq:pot}, \eqref{eq:tranpot} is equivalent to
\ba
\label{eq:pot1}
-\nabla \cdot (\epsilon_1 \nabla \psi_1) & = & q_1, \quad x \in \Omega_1; \quad
\psi_1|_\Gamma = \lambda \\
\label{eq:pot2}
-\nabla \cdot (\epsilon_2 \nabla \psi_2) & = & q_2, \quad x \in
\Omega_2; \quad \psi_2|_\Gamma = \lambda + \psi_\triangle,
\ea
which requires $\left[\epsilon \dpd{\psi}{\nu}
  \right]_\Gamma = 0$.  The algorithm DDP we proposed in \cite{haii}
is essentially a modification of the Neumann-Neumann algorithm
\cite{qvddbook}.

\begin{shaded}
\emph{\textbf{Algorithm DDP to solve \eqref{eq:pot}, \eqref{eq:tranpot}}}:
Given $\lambda^{(0)}$, for each $k\geq 0$,
\begin{enumerate}[leftmargin=.5in]
\item Solve \eqref{eq:pot1} and \eqref{eq:pot2} for $\psi_i^{(k+1)}$, $i =1,2$.
\item Update $\lambda$ by
\bas
\lambda^{(k+1)} = \lambda^{(k)} - \theta_{\psi}
\left[\epsilon \dpd{\psi^{(k+1)}}{\nu}\right]_\Gamma
\eas
\item Continue with (1) unless stopping criterium
$\left\| \left[\epsilon \dpd{\psi^{(k+1)}}{\nu} \right]_\Gamma \right\|$ holds.
\end{enumerate}
\end{shaded}

\subsection{Domain Decomposition for Continuity Equation \eqref{eq:nsb},
\eqref{eq:tranu}}
\label{sub:contalg}

Here we seek to find data $\lambda$ so that \eqref{eq:nsb},
\eqref{eq:tranu} is equivalent to the problem:
\ba
\label{eq:u1}
-\nabla \cdot (D_{n_1} \delta_n^2 e^{\psi_1} \nabla u_1) & = & R_1, \quad x \in \Omega_1;
\quad
u_1|_\Gamma = \lambda \\
\label{eq:u2}
-\nabla \cdot (D_{n_2} \delta_n^2 e^{\psi_2} \nabla u_2) & = & R_2, \quad x \in \Omega_2;
\quad \\
\label{eq:algtran}
& & \quad u_2|_\Gamma =
\frac{a_1^n (e^{\psi})_1^\Gamma}{a_2^n (e^{\psi})_2^\Gamma}\lambda
+ \frac{J_{n_1}}{a_2^n \delta^2 (e^{\psi})_2^\Gamma}
\ea
which requires the homogeneous jump condition $\left[J_n\right]_\Gamma
= 0$.

Algorithm {\bf DDC} proposed in this paper is very different from
\textbf{DDP} because it proceeds sequentially from domain $\Omega_1$
to domain $\Omega_2$. In addition, while it corrects $\lambda$ in a
manner similar to a Neumann-Neumann algorithm, in \eqref{eq:algtran}
it takes advantage of Neumann data from $\Omega_1$ resulting from
\eqref{eq:u1}. An appropriate parallel algorithm for {\bf DDC} which
uses Neumann rather than Dirichlet data as in \eqref{eq:u1},
\eqref{eq:u2} was promising for a synthetic example, but it has
difficulties with convergence for realistic devices. \myskip{We are
  studying other possible extensions.}

\begin{shaded}
\emph{\textbf{Algorithm DDC
to solve \eqref{eq:nsb}, \eqref{eq:tranu} or \eqref{eq:psb}, \eqref{eq:tranv}}}:
Given $\lambda^{(0)}$, for each $k\geq 0$,
\begin{enumerate}[leftmargin=.5in]
\item Solve \eqref{eq:u1} for $u_1^{(k+1)}$ and then solve
  \eqref{eq:u2}--\eqref{eq:algtran} for $u_2^{(k+1)}$.
\item Update $\lambda$ by
\bas
\lambda^{(k+1)} = \lambda^{(k)} - \theta_n
\left[D_n e^{\psi} \dpd{u^{(k+1)}}{\nu}\right]_\Gamma
\eas
\item Continue with (1) unless stopping criterium
$\left\| \left[J_n\right]_\Gamma \right\|$ holds.
\end{enumerate}
\end{shaded}

While DDP and DDC are motivated by the multiphysics nature of the
model, they may be viewed as extensions of Neumann-Neumann iterative
substructuring methods to nonhomogeneous jumps and Robin-like
transmission conditions. A scalable parallel implementation may be
achieved in the future using two-level techniques [\cite{qvddbook}
  \textsection 3.3.2].

\section{Heterojunction Semiconductor Simulation}
\label{sec:scsim}

Now we present numerical simulation results.  Structure 1 is synthetic
and solar cell-like, and is made of two hypothetical materials we call
L1 and R1. Structure 2 is made of Si and Cu$_3$PSe$_4$ (CPS). In
Table~\ref{tab:matprop} we give details.

\begin{figure}
\centering
\includegraphics[width=.75\textwidth]{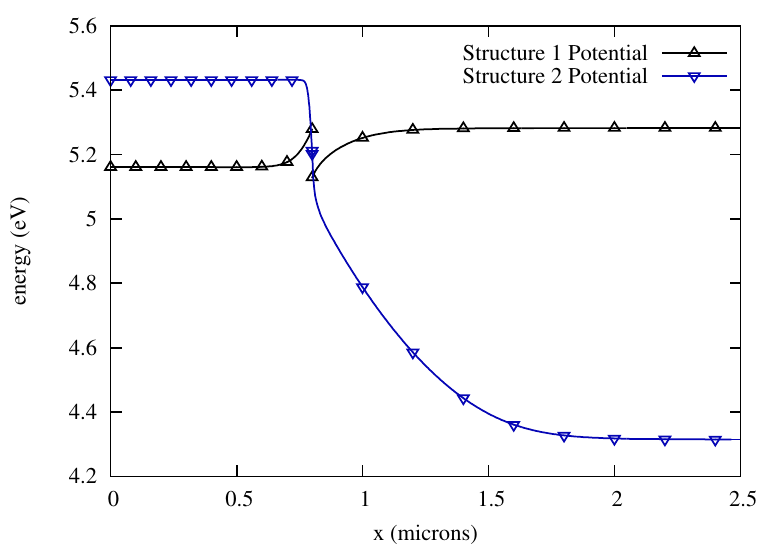}\\
\includegraphics[width=.75\textwidth]{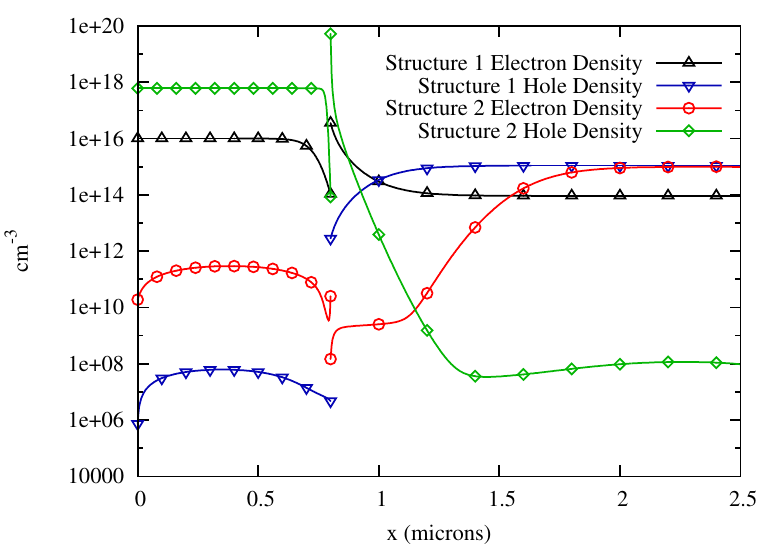}\\
\includegraphics[width=.75\textwidth]{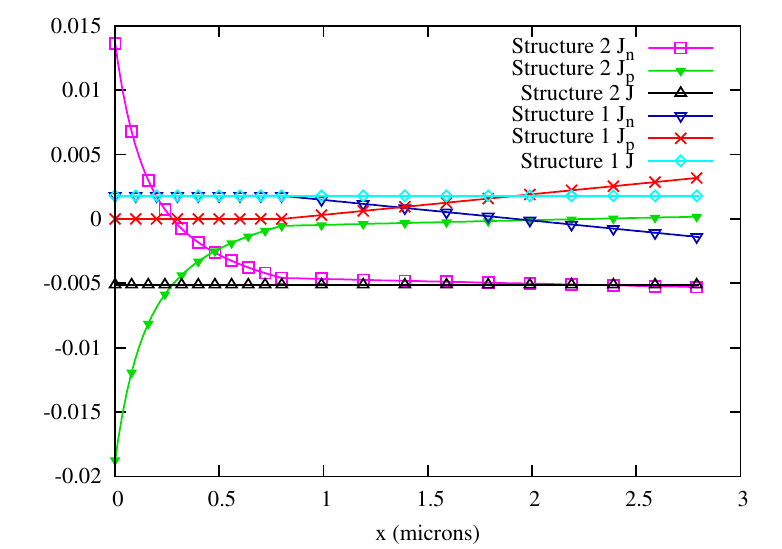}
\caption{\label{fig:mondd}Simulation of Structure 1 and 2 with DDM}
\end{figure}

We use DFT to calculate $\psi_\triangle = -0.01$ eV for the
Cu$_{0.75}$P$_{0.25}$-Si interface formed from CPS (001) and the Si
(111) surfaces having normally oriented dangling bonds.
Next we apply Domain Decomposition and specifically the algorithms
{\bf DDP, DDC}; see Figure~\ref{fig:mondd}.
For both structures we see the impact of heterojunction and large
jumps of $\psi,n,p$ across the interface. The results are validated
with a hard-coded monolithic solver.

As concerns solver's performance, in Table~\ref{tab:meshind} we show
that \textbf{DDC} is mesh independent, similarly to \textbf{DDP}
\cite{haii}. \myskip{Thus, the complexity of the DDM is comparable to
  that of monolithic solver.} Furthermore, the choice of $\theta$ is
crucial. (In forthcoming work \cite{haii3} we show how $\theta$ is determined from
analysis of the jump data.)

\begin{table}[th!]
\begin{shaded}
\centering
\scriptsize
\caption{\label{tab:meshind}Number of iterations at each Gummel
    Iteration (GI) and parameters $\theta_n,\theta_p$
      for Structure 1 and algorithm \textbf{DDC}. \textbf{DDP} uses
      $\theta^1_{\psi} = 0.0025$, $\theta_{\psi}^2 = 0.00025$. Also,
      we use $\theta^2_n=4e11$, $\theta^2_p=1.4$}
  \begin{tabular}{l | c c c c | c c c c}
    & & DDC u, & $\theta^1_n = 2.5$ & & & DDC v, & $\theta^1_p = 180$ & \\
    \hline
    N & GI 1 & GI 2 & GI 3 & GI 4 & GI 1 & GI 2 & GI 3 & GI 4 \\
    \hline
    201 & 6 & 2 & 1 & 1 & 5 & 3 & 1 & 1 \\
    401 & 5 & 2 & 1 & 1 & 8 & 4 & 1 & 1 \\
    601 & 3 & 2 & 1 & 1 & 8 & 4 & 1 & 1 \\
    801 & 4 & 2 & 1 & 1 & 8 & 4 & 1 & 1 \\
\hline
  \end{tabular}
  \end{shaded}
\end{table}

\begin{table}[th!]
\begin{shaded}
\centering
\scriptsize
  \caption{\label{tab:matprop}Material and Structure Parameters}
  \begin{tabular}{lrrrr}
    property & L1 & R1 & CPS & Si \\
    \hline
permittivity $\epsilon$ & 10.0 & 10.0 & 15.1 \cite{Foster} & 11.9  \cite{Sze} \\
electron affinity $\chi$ (eV)& 5 & 5 & 4.05 & 4.05 \cite{Sze} \\
band gap $E_g$ (eV) & 1.0 & 0.5 & 1.4 \cite{Foster} & 1.12 \cite{Sze} \\
eff.~electron density of states $\tilde{N}_C$ (cm$^{-3}$) & $5\times10^{18}$
& $5\times10^{18}$ & $3\times10^{19}$ & $2.8\times10^{19}$ \\
eff.~hole density of states $\tilde{N}_V$ (cm$^{-3}$) & $5\times10^{18}$
& $5\times10^{18}$ & $1.2\times10^{18}$ & $1\times10^{19}$ \\
dopant charge density $\tilde{N}_T$ (cm$^{-3}$) & $1\times10^{16}$
& $-1\times10^{15}$ & $-6\times10^{17}$ \cite{Foster} & $1\times10^{15}$ \\
electron diffusion constant $\tilde{D}_n$ (cm$^2$/s) & 2.0 & 2.0
& 2.6  & 37.6 \cite{Sze} \\
hole diffusion constant $\tilde{D}_p$ (cm$^2$/s) & 1.0 & 1.0 & 0.5 & 12.9 \cite{Sze} \\
constant photogeneration density $G$ (cm$^{-3}$/s) & $1\times10^{17}$
& $1\times10^{20}$ & $1\times10^{21}$ & $1\times10^{18}$ \\
direct recombination constant $R_{dc}$ (cm$^3$/s) & $1\times10^{-10}$
& $1\times10^{-10}$ & $1\times10^{-10}$ & $1\times10^{-15}$ \\
jump in potential $\psi_\triangle$ (eV) & \multicolumn{2}{c}{-0.15}
& \multicolumn{2}{c}{-0.01} \\
%    ($A_n T^2$)(A cm$^2$) & \multicolumn{2}{c}{} & \multicolumn{2}{c}{} \\
\hline
  \end{tabular}
\end{shaded}
\end{table}

\section{Conclusions}
\label{sec:conc}

The main contribution reported in this paper
advances HPC methodology for solving
  problems with complex interface physics.
We presented DDM for the simulation of charge transport in
heterojunction semiconductors. Our method allows the coupling of
``black-box'' D-D (drift diffusion) solvers in subdomains
corresponding to single semiconductor materials. We compared DDM to a
monolithic solver, and the results are promising; see Table~
\ref{tab:timeandqoicomparison}. As usual, DDM approach wins
  for large $N$. Also, it works when monolithic solver fails.
In the model presented here DFT is used to determine heterojunction
parameters but is currently entirely decoupled from D-D solvers in the
bulk subdomains.  Our approach is a first step towards a true
multiscale simulation coupling the atomic and device scales.

\begin{table}[th!]
\begin{shaded}
\centering
\scriptsize
  \caption{\label{tab:timeandqoicomparison}Efficiency of DDM vs
    monolithic solvers. Column 4 estimates multicore efficiency.}
  \begin{tabular}{l | c c c c}
    $N$ & Monolithic time (sec) & DDM time (sec)
      & DDM parallel estimate & Current \\
    \hline
    501/501 & 0.5494 & 1.313 & 0.8 & 0.006637 \\
    751/751 & 0.8122 & 1.4537 & 0.9 & 0.006649\\
    1001/1001 & 1.0231 & 1.4173 & 0.9 & 0.006655 \\
    1251/1251 & failed & 2.1665 & 1.3 & 0.0066592 \\
\hline
  \end{tabular}
\end{shaded}
\end{table}

At the current stage, the computational complexities of the
  microscale and macroscale simulations are vastly different.  The
  microscale DFT simulations using VASP solver \cite{kressel} for
  electronic structure simulations running on 4 machines with 12 cores
  with MPI2, take several days to complete.
 On the other hand, the D-D solver takes less than minutes at worst to
 complete; see Table~\ref{tab:timeandqoicomparison}. Thus, a true
coupled multiscale approach is not feasible yet.

More broadly, problems with nonhomogeneous jump conditions
  across interfaces only begin to be investigated from mathematical
  and computational point of view. Our DDM approach is a new paradigm
  that applies elsewhere, e.g., for discrete fracture approximation
  models where nonhomogeneous jump conditions arise
  \cite{frih,martin2005modeling}.

\textbf{Acknowledgements:} This research was partially supported by
the grant NSF-DMS 1035513 grant ``SOLAR: Enhanced Photovoltaic
Efficiency through Heterojunction assisted Impact Ionization.'' We
would like to thank G.~Schneider for useful discussions.

\bibliographystyle{amsplain}
%\bibliography{haii2.bib}
\providecommand{\bysame}{\leavevmode\hbox to3em{\hrulefill}\thinspace}
\providecommand{\MR}{\relax\ifhmode\unskip\space\fi MR }
\providecommand{\MRhref}[2]{%
  \href{http://www.ams.org/mathscinet-getitem?mr=#1}{#2}
}
\providecommand{\href}[2]{#2}

\end{document}